\documentclass[preprint,preprintnumbers,showpacs,showkeys,amsmath,amssymb,nofootinbib,superscriptaddress]{revtex4}

\usepackage[dvips]{graphicx}
\usepackage{color}
\usepackage{latexsym}
\usepackage{amsmath}
\usepackage{amssymb}
\usepackage{eufrak}
\usepackage{euscript}
\usepackage{pstricks}
\usepackage{graphics}
\usepackage{graphicx}
\usepackage{picture}
\def\[{\left\lbrack}
\def\]{\right\rbrack}

\def\({\left(}
\def\){\right)}

\newcommand{\bee}{\begin{equation}}
\newcommand{\eee}{\end{equation}}
\newcommand{\eaa}{\end{eqnarray}}
\newcommand{\baa}{\begin{eqnarray}}

\def\ni{\noindent}

\begin{document}

\title{\Large Holographic considerations on non-gaussian statistics and gravothermal catastrophe}


\author{Everton M. C. Abreu}\email{evertonabreu@ufrrj.br}
\affiliation{Grupo de F\' isica Te\'orica e Matem\'atica F\' isica, Departamento de F\'{i}sica, Universidade Federal Rural do Rio de Janeiro, 23890-971, Serop\'edica, RJ, Brazil}
\affiliation{Departamento de F\'{i}sica, Universidade Federal de Juiz de Fora, 36036-330, Juiz de Fora, MG, Brazil}
\author{Jorge Ananias Neto}\email{jorge@fisica.ufjf.br}
\affiliation{Departamento de F\'{i}sica, Universidade Federal de Juiz de Fora, 36036-330, Juiz de Fora, MG, Brazil}
\author{Edesio M. Barboza Jr.} \email{edesiobarboza@uern.br}
\affiliation{Departamento de F\'isica, Universidade do Estado do Rio Grande do Norte, 59610-210, Mossor\'o, RN, Brazil}
\author{Rafael da C. Nunes}\email{rafaelda.costa@e-campus.uab.cat}
\affiliation{Departmento de F\' isica, Universitat Aut\'onoma de Barcelona, 08193, Bellaterra, Barcelona, Spain}
\affiliation{Funda\c{c}\~ao CAPES, Minist\'erio da Educa\c{c}\~ao e Cultura, 70040-020, Bras\' ilia, DF, Brazil}

\date{\today}





\begin{abstract}
\noindent In this paper we have derived the equipartition law of energy using Tsallis formalism and the Kaniadakis power law statistics in order to obtain a modified gravitational constant. We have applied this result in the gravothermal collapse phenomenon. We have discussed the equivalence between Tsallis and the Kaniadakis statistics in the context of Verlinde entropic formalism.  
In the same way we have analyzed negative heat capacities in the light of gravothermal catastrophe.
The relative deviations of the modified gravitational constants are derived. 
\end{abstract}

\maketitle

\section{Introduction}

The mechanism of gravothermal instability, discovered by Antonov \cite{antonov,lw,padmanabhan,chavanis} is an important phenomena in gravitational thermodynamics.   It has been very helpful for an extensive research concerning statistical mechanics of long range interactions systems in several fields in physics \cite{agr}.   This connection with thermodynamics and statistical mechanics has motivated us to investigate statistically the gravothermal catastrophe

At the same time, there are theoretical evidences that the understanding of gravity has been greatly benefited from a possible connection with thermodynamics. Pioneering works of Bekenstein  \cite{BEK} and Hawking  \cite{HAW} have described this issue. For example, quantities as area and mass of black-holes are associated with entropy and temperature respectively. Working on this subject, Jacobson  \cite{Jac} interpreted Einstein field equations as a thermodynamic identity. Padmanabhan  \cite{PAD} gave an interpretation of gravity as an equipartition theorem. 

Recently, Verlinde  \cite{verlinde} brought an heuristic derivation of gravity, both Newtonian and relativistic, at least for static spacetime.  The equipartition law of energy has also played an important role. 
On the other hand, one can ask: what is the contribution of gravitational models in thermodynamics theories?  

The concept introduced by Verlinde is analogous to Jacobson's  \cite{Jac}, who proposed a thermodynamic derivation of Einstein's equations.  The result has shown that the gravitation law derived by Newton can be interpreted as an entropic force originated by perturbations in the information ``manifold" caused by the motion of a massive body when it moves 
away from the holographic screen.  Verlinde used this idea  together with the Unruh result  \cite{unruh} and he obtained Newton's second law.   The entropic force together with the holographic principle and the equipartition law of energy resulted in Newton's law of gravitation.  Besides, Verlinde's ideas have been used in cosmology  \cite{abreu}.

Currently, two of the most investigated extensions of the usual Boltzmann-Gibbs (BG) theory  are  Tsallis thermostatistics theory  \cite{tsallis,tsallis2,tsallis3} (TT) and Kaniadakis power law statistics  \cite{kani1,kani2,kani3} (KS). The former case, TT, initially considers the entropy formula as a nonextensive (NE) quantity where there is a parameter $q$ that measures the so-called degree of nonextensivity. This formalism has been successfully applied in many physical models.  An important feature is that when $q\rightarrow1$ we recover the usual BG theory. On the other hand, the so called KS naturally emerges from the context of special relativity and in the kinetic interaction principle. This formalism has also been successfully applied in many physical models. There is a parameter $\kappa$ in the KS that in the limit $\kappa\rightarrow0$ the BG theory is also recovered.

This paper has two parts: in the first one, the self-gravitating system  \cite{SA} (and references therein), is discussed in the context of the TT and KS formalisms. We found that for both formalisms the dependence of the specific heat, $C_V$, on the nonextensive parameter $q$ and on the $\kappa$ parameter give rise to a negative branch, a result that features the gravothermal collapse. In the second part, we have used the TT formalism and the KS in the Verlinde formalism  \cite{verlinde}. As a result, the equipartition theorem is derived in the framework of KS leading to a modified gravitational constant.

This paper is organized in as follows: In section \ref{TTS} the main steps of Tsallis, Kaniadakis and Verlinde's formalisms are reviewed. 
In section \ref{mgc} a modified gravitational constant is obtained in the light of Kaniadakis' formalism.  In Section \ref{alt-way}, in an holographic background, we have obtained relations between $q$ and $\kappa$.  In section \ref{GC} the gravothermal collapse is studied in the light of TT and KS formalisms.  
In section \ref{pm2} the relative errors of the gravitational constants are calculated . The conclusions and perspectives are presented in the last section.

\section{Nonextensive statistics and holographic entropy}
\label{TTS}
\renewcommand{\theequation}{2.\arabic{equation}}
\setcounter{equation}{0}

The study of entropy has been an interesting task through recent years thanks to the fact that it can be understood as a measure of information loss concerning the microscopic degrees of freedom of a physical system, when describing it in terms of macroscopic variables.  Appearing in different scenarios, we can conclude that entropy can be considered as a consequence of the gravitational framework \cite{BEK,HAW}.
These issues motivated some of us to consider other alternatives to the standard BG theory in order to work with Verlinde's ideas together with other subjects\cite{abreu}.

The objective of this section is to provide the reader with the main tools that will be used in the following sections.   Although both formalisms are well known in the literature, these brief reviews can emphasize precisely that there is a connection between both ideas which was established recently  \cite{ananias}.

\subsection{Tsallis' formalism}

Tsallis \cite{tsallis} has proposed an important extension of the BG statistical theory and curiously, in a technical terminology, this model is also currently referred to as nonextensive statistical mechanics. TT formalism defines a nonadditive entropy given by

\begin{eqnarray}
\label{nes}
S_q =  k_B \, \frac{1 - \sum_{i=1}^W p_i^q}{q-1}\;\;\;\;\;\;\qquad \Big(\,\sum_{i=1}^W p_i = 1\,\Big)\,\,,
\end{eqnarray}

\ni where $p_i$ is the probability of the system to be in a microstate, $W$ is the total number of configurations and 
$q$, known in the current literature as Tsallis parameter or NE  parameter, is a real parameter which quantifies the degree of nonextensivity. 
The definition of entropy in TT formalism possesses the usual properties of positivity, equiprobability, concavity and irreversibility and motivated the study of multifractals systems.
It is important to stress that Tsallis formalism contains the BG statistics as a particular case in the limit $ q \rightarrow 1$ where the usual additivity of entropy is recovered. Plastino and Lima \cite{PL} 
have derived a NE equipartition law of energy. It has been shown that the kinetic foundations of Tsallis' NE statistics lead to a velocity distribution for free particles given by  \cite{SPL}

\begin{eqnarray}
\label{vd}
f_q(v)=B_q \[1-(1-q) \frac{m v^2}{2 k_B T}\]^{1/1-q},
\end{eqnarray}
where $B_q$ is a normalization constant. The expectation value of $v^2$ is given by \cite{SA}

\begin{eqnarray}
\label{v2}
<v^2>_q=\frac{\int^\infty_0\, f_q \,v^2 dv}{\int^\infty_0\, f_q dv}\\ \nonumber
=\frac{2}{5-3q}\, \frac{k_B T}{m}.
\end{eqnarray}
The equipartition theorem is then obtained by using 

\begin{eqnarray}
\label{reqq}
E_q=\frac{1}{2} N m <v^2>_q,
\end{eqnarray}
where we arrive at
\begin{eqnarray}
\label{ge}
E_q = \frac{1}{5 - 3 q} N k_B T\,\,.
\end{eqnarray}
The range of $q$ is $ 0 \le q < 5/3 $.  For $ q=5/3$ (critical value) the expression of the equipartition law of energy, Eq. (\ref{ge}), diverges. It is also easy to observe that for $ q = 1$,  the classical equipartition theorem for each microscopic degrees of freedom can be recovered.

\subsection{Kaniadakis' statistics}

Kaniadakis statistics \cite{kani1}, also called $\kappa$-statistics, similarly to TT formalism generalizes the standard BG statistics initially by the introduction of $\kappa$-exponential and $\kappa$-logarithm defined by

\begin{eqnarray}
\label{expk}
exp_\kappa(f)=\( \sqrt{1+\kappa^2 f^2}+\kappa f \)^\frac{1}{\kappa},
\end{eqnarray}

\begin{eqnarray}
\label{logk}
\ln_\kappa(f)=\frac{f^\kappa-f^{-\kappa}}{2\kappa},
\end{eqnarray}
with the following operation being satisfied
\begin{eqnarray}
\ln_\kappa\(exp_\kappa(f)\)=exp_\kappa\(\ln_\kappa(f)\)\equiv f.
\end{eqnarray}
By Eqs. (\ref{expk}) and (\ref{logk}) we can observe that the $\kappa$-parameter deforms the usual definitions of the exponential and logarithm functions.

The $\kappa$-entropy associated with this $\kappa$-framework is given by
\begin{eqnarray}
S_\kappa(f)=-\int d^3 p \,f \[\frac{f^\kappa-f^{-\kappa}}{2\kappa}\],
\end{eqnarray}
which recovers the BG entropy in the limit $\kappa \rightarrow 0$. It is important to mention here that the $\kappa$-entropy satisfied the properties of concavity, additivity and extensivity. Tsallis' entropy satisfies the property of concavity and extensivity but not additivity. This property is not fundamental, in principle. The $\kappa$-statistics has been successfully applied in many experimental fronts. As an example we can mention cosmic rays \cite{Kanisca1}, quark-gluon plasma \cite{Tewe}, kinetic models describing a gas of interacting atoms and photons \cite{Ross} and financial models \cite{RBJ}.

The kinetic foundations of $\kappa$-statistics lead to a velocity distribution for free particles given by \cite{BSS}

\begin{eqnarray}
f_\kappa(v)=\[ \sqrt{1+\kappa^2 \( -\frac{m v^2}{2 k_B T}\)^2}  - \frac{\kappa m v^2}{2 k_B T} \]^\frac{1}{\kappa}.
\end{eqnarray}
The expectation value of $v^2$ is given by
\begin{eqnarray}
\label{vk}
<v^2>_\kappa=\frac{\int^\infty_0\, f_\kappa \,v^2 dv}{\int^\infty_0\, f_\kappa dv}.
\end{eqnarray}
Using the integral relation \cite{kani2}

\bee
\int_0^\infty x^{r-1}\, exp_\kappa (-x) dx
\,=\,\frac{[1+(r-2)\left|\kappa\right|] \left|2\kappa\right|^{-r}]}{[1-(r-1)\left|\kappa\right| \,]^2-\kappa^2}\; \frac{\Gamma\(\frac{1}{\left|2\kappa\right|}-\frac{r}{2}\)}{\Gamma\(\frac{1}{\left|2\kappa\right|}+\frac{r}{2}\)}\; \Gamma(r),
\eee
we have

\begin{eqnarray}
\label{vk2}
<v^2>_\kappa=\,\frac{(1-\frac{1}{2} \kappa)^2\,(1+\frac{3}{2} \kappa)}{(1-\frac{3}{2} \kappa)^2\,(1+\frac{1}{2} k) \, 2\kappa}\; \;
\frac{\Gamma{(\frac{1}{2\kappa}-\frac{3}{4})\,\Gamma{(\frac{1}{2\kappa}+\frac{1}{4})}}}{\Gamma{(\frac{1}{2\kappa}+\frac{3}{4})\,\Gamma{(\frac{1}{2\kappa}-\frac{1}{4})}}}\; \frac{k_B T}{m}.\\ \nonumber
\end{eqnarray}
The $\kappa$-equipartition theorem is then obtained by using

\begin{eqnarray}
\label{reqk}
E_\kappa=\frac{1}{2} N m <v^2>_\kappa,
\end{eqnarray}
where we arrive at

\begin{eqnarray}
\label{keq}
E_\kappa=\frac{1}{2} N \;\; \frac{(1-\frac{1}{2} \kappa)^2\,(1+\frac{3}{2} \kappa)}{(1-\frac{3}{2} \kappa)^2\,(1+\frac{1}{2} k) \, 2\kappa}\; \;
\frac{\Gamma{(\frac{1}{2\kappa}-\frac{3}{4})\,\Gamma{(\frac{1}{2\kappa}+\frac{1}{4})}}}{\Gamma{(\frac{1}{2\kappa}+\frac{3}{4})\,\Gamma{(\frac{1}{2\kappa}-\frac{1}{4})}}}\;\;\; k_B T.
\end{eqnarray}
The range of $\kappa$ is $ 0 \le \kappa < 2/3 $.  For $ \kappa=2/3$ (critical value) the expression of the equipartition law of energy, Eq. (\ref{keq}), diverges. For $ \kappa = 0$, the classical equipartition theorem for each microscopic degrees of freedom can be recovered. 

\subsection{Verlinde's Formalism}

The formalism proposed by E. Verlinde  \cite{verlinde} obtains the gravitational acceleration  by using the holographic principle and the well known equipartition law of energy. His ideas relied on the fact that gravitation can be considered universal and independent of the details of the spacetime microstructure.  Besides, he brought new concepts concerning holography since the holographic principle must unify matter, gravity and quantum mechanics.

The model considers a spherical surface as being the holographic screen, with a particle of mass $M$ positioned in its center. The holographic screen can be imagined as a storage device for information. The number of bits, which is the smallest unit of information in the holographic screen, is assumed to be proportional to the  holographic screen
area $A$
\begin{eqnarray}
\label{bits}
N = \frac{A }{l_P^2},
\end{eqnarray}
where $ A = 4 \pi r^2 $ and $l_P = \sqrt{\frac{G\hbar}{c^3}}$ is the Planck length and $l_P^2$ is the Planck area.   In Verlinde's framework one can suppose that the bits total energy on the screen is given by the equipartition law of energy

\begin{eqnarray}
\label{eq}
E = \frac{1}{2}\,N k_B T.
\end{eqnarray}

\ni It is important to notice that the usual equipartition theorem in Eq. (\ref{eq}), can be derived from the usual BG thermostatistics. 
Let us consider that the energy of the particle inside the holographic screen is equally divided by all bits in such a manner that we can have the expression

\begin{eqnarray}
\label{meq}
M c^2 = \frac{1}{2}\,N k_B T.
\end{eqnarray}

\ni With Eq. (\ref{bits}) and using the Unruh temperature equation  \cite{unruh} given by

\begin{eqnarray}
\label{un}
k_B T = \frac{1}{2\pi}\, \frac{\hbar a}{c},
\end{eqnarray}

\ni we are  able to obtain the  (absolute) gravitational acceleration formula

\begin{eqnarray}
\label{acc}
a &=&  \frac{l_P^2 c^3}{\hbar} \, \frac{ M}{r^2}\nonumber\\ 
&=& G \, \frac{ M}{r^2}\,\,.
\end{eqnarray}
From Eq. (\ref{acc}) we can see that the Newton constant $G$ is just written in terms of the fundamental constants, $G=l_P^2 c^3/\hbar$.

\section{Modified gravitational Constant}
\label{mgc}
\renewcommand{\theequation}{3.\arabic{equation}}
\setcounter{equation}{0}

As an application of NE equipartition theorem in Verlinde's formalism we can
use the NE equipartition formula, i.e., Eq. (\ref{ge}).  Hence, we can obtain a modified acceleration formula given by  \cite{abreu}
\begin{eqnarray}
\label{accm}
a = G_q \, \frac{ M}{r^2},
\end{eqnarray}
where $G_q$ is an effective gravitational constant which is written as

\bee
\label{S}
G_q=\,\frac{5-3q}{2}\,G\,\,.
\eee
From result (\ref{S}) we can observe that the effective gravitational constant depends on the NE parameter $q$. For example, when $q=1$ we have $ G_q=G$ (BG scenario) and for $q\,=\,5 / 3$ we have the curious and hypothetical result which is $G_q=0$.  This result shows us that $q\,=\,5/3$ is an upper bound limit when we are dealing with the holographic screen.  Notice that this approach is different from the one demonstrated in  \cite{cn}, where the authors considered in their model that the number of states is proportional to the volume and not to the area of the holographic screen.

On the other hand, if we use the Kaniadakis equipartition theorem, Eq.(\ref{keq}), in the Verlinde formalism, the modified acceleration formula is given by
\begin{eqnarray}
\label{acck}
a = G_{\kappa} \, \frac{ M}{r^2}\,\,,
\end{eqnarray}

\ni where $G_{\kappa}$ is an effective gravitational constant which is written as

\begin{eqnarray}
\label{Gk}
G_{\kappa}=  \frac{(1-\frac{3}{2} \kappa)^2\,(1+\frac{1}{2} \kappa) \, 2\kappa}{(1-\frac{1}{2} \kappa)^2\,(1+\frac{3}{2} \kappa)}\; \;
\frac{\Gamma{(\frac{1}{2\kappa}+\frac{3}{4})\,\Gamma{(\frac{1}{2\kappa}-\frac{1}{4})}}}{\Gamma{(\frac{1}{2\kappa}-\frac{3}{4})\,\Gamma{(\frac{1}{2\kappa}+\frac{1}{4})}} }\; \; G\,\,,
\end{eqnarray}

\ni which can be written alternatively as $G=f(\kappa)\,G_{\kappa}$ used in Eq. (4.10)-(4.13), as we have explained in Section IV.

From result (\ref{Gk}) we can observe that the effective gravitational constant depends on the $\kappa$ parameter. For example, using the Gamma functions properties written in Eq. (\ref{cc2}) 
we obtain for $\kappa=0$ that $G_{\kappa}=G$ (BG scenario). For $\kappa\,=\,2 / 3$ and taking into account that the only Gamma function that diverge for this value ($\Gamma{(\frac{1}{2\kappa}-\frac{3}{4}})$) is in the denominator of Eq.(\ref{Gk}), 
we have a result which is $G_\kappa=0$.

From the limits $\kappa=0, q=1$, which we recover the BG statistics, and $\kappa=2/3, q=5/3$, we can established a linear relation between $\kappa$ and $q$ written as

\begin{eqnarray}
\label{rel}
\kappa=q-1.
\end{eqnarray}
In particular, the limits $\kappa=0, q=1$ and $\kappa=\frac{2}{3}, q=\frac{5}{3}$ lead to $G_{q}=G_{\kappa}$. It is clear that relation $(\ref{rel})$ is valid only for the range $1\leq q < 5/3$.
In Figure 2 we have plotted $G_{q}/G$, Eq.(\ref{S}), and $G_\kappa/G$, Eq.(\ref{Gk}), as function of $\kappa$-parameter.    The dashed line represents the Tsallis formalism and the solid line represents the Kaniadakis formalism.

\begin{figure}[ih]
\includegraphics[scale=1.1,angle=0]{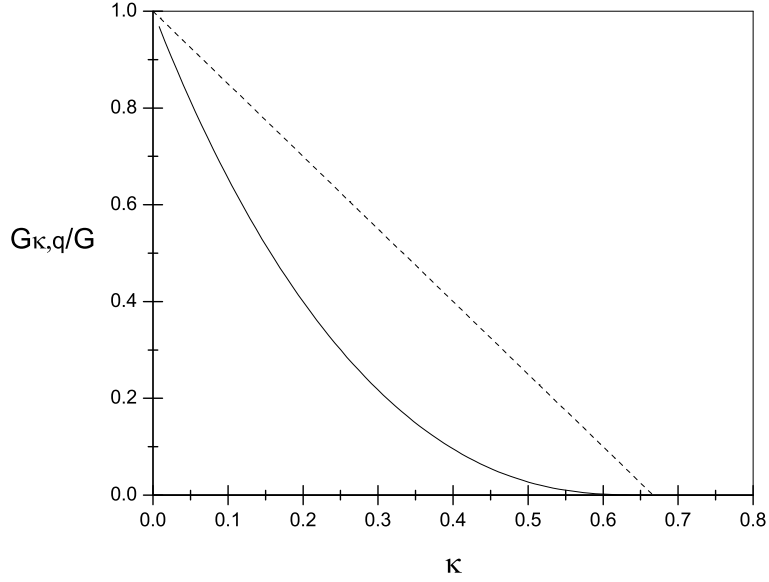}
\caption{The normalized modified gravitational constants for Tsallis (dashed line$=G_q /G$) and Kaniadakis (solid line$=G_{\kappa}/G$) formalisms as functions of $\kappa$.Note that $G_q$ is closer to $G$ than $G_{\kappa}$. Also note that the strength for gravitational field is greater in TT formalism than in the Kappa one.}
\end{figure}

In order to explicit the difference between the gravitational constants obtained by Tsallis, Eq. (\ref{S}), and Kaniadakis (Eq. (\ref{Gk})) formalisms, we define a new function as 

\begin{eqnarray}
\label{dg}
DG=\frac{G_q-G_\kappa}{G}.
\end{eqnarray}
In Figure 3 we plot $DG$ as a function of the deformation parameter $\kappa$.

\begin{figure}[t]
\includegraphics[scale=1.1,angle=0]{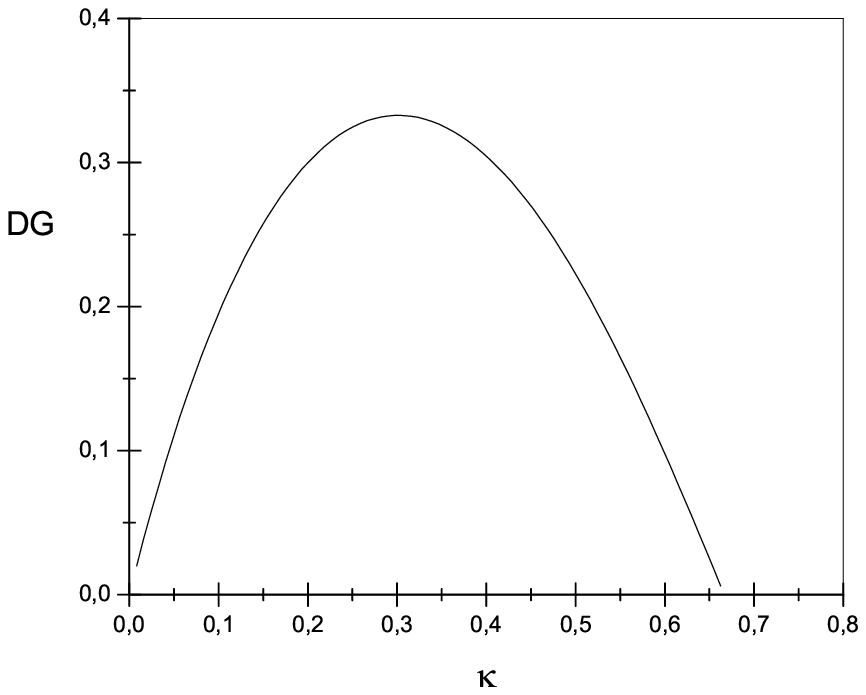}
\caption{The normalized difference between modified gravitational constants obtained by Tsallis and Kaniadakis statistics, Eq.(\ref{dg}), as a function of $\kappa$.}
\end{figure}

\section{An algebraic alternative  way of obtaining the relationship between $q$ and $k$}
\label{alt-way}
\renewcommand{\theequation}{4.\arabic{equation}}
\setcounter{equation}{0}

\bigskip


We know that to establish values and/or intervals for $q$ is important in different physical scenarios, as described in the  last sections (and in the quoted references).  Hence, we believe that it would be interesting to obtain $q$ -relations.  Having said that let us make an assumption and test its validity. 
In other words, let us explain that the relations obtained from now on will be obtained in the entropic (Verlinde) scenario explored and it is now a general result since TT and KS are very different formalisms.  So, what we will see now can be realized as an algebraic conjecture and not an equivalence between both formalisms.  These words have to be well understood.

From (\ref{S}) and (\ref{Gk}) let us assume that

\bee
\label{cc1}
\frac{2}{5-3q}\,=\,\frac{(1-\frac{1}{2} \kappa)^2\,(1+\frac{3}{2} \kappa)}{(1-\frac{3}{2} \kappa)^2\,(1+\frac{1}{2} k) \, 2\kappa}\; \;
\frac{\Gamma{(\frac{1}{2\kappa}-\frac{3}{4})\,\Gamma{(\frac{1}{2\kappa}+\frac{1}{4})}}}{\Gamma{(\frac{1}{2\kappa}+\frac{3}{4})\,\Gamma{(\frac{1}{2\kappa}-\frac{1}{4})}}}\,,
\eee

\ni and for $k \longrightarrow 0$ we must have $q \longrightarrow 1$.   Using Eq. (\ref{cc2}) we have that

\bee
\label{cc3}
\frac{2}{5-3q}\,=\,\frac{\Big(1-\frac 12 k\Big)^2\Big(1+\frac 32 k\Big)}{\Big(1-\frac 32 k\Big)^2\Big(1+\frac 12 k\Big)}.
\eee

\ni Since $k$ is very small ($k \longrightarrow 0$), let us eliminate $k~3$-terms.  So, from (\ref{cc3}) we have that

\bee
\label{cc4}
\Big[ \frac{3}{2(5-3q)} \,+\,\frac 54 \Big]\,k^2\,-\, \Big[ \frac{5}{5-3q}\,+\,\frac 12 \Big] k\,+\, \frac{2}{5-3q}\,-\,1\,=\,0,
\eee

\ni and using $k=0$ in (\ref{cc4}) we have that $q=1$.  From (\ref{cc3}) again, if we eliminate $k^2$ and $k^3$-terms, we have that

\bee
\label{cc5}
1\,+\,\frac 12 k \,=\, \frac {2}{5-3q} \Big[1\,-\,\frac 52 k \Big],
\eee

\ni so $k=0  \Longrightarrow q=1$.

From (\ref{cc4}), let us invert the order , i.e., let us make $q=1$ in (\ref{cc4}) we have that $k^2\,-\,\frac 32 k\,=\,0 \Longrightarrow k_1 = 0$ and $k_2 = 3/2$.  And the root that matters is $k=0$.  From (\ref{cc3}) we have that

\bee
\label{cc6}
k^3\,+\, \frac 23 \Big(\frac{31-15q}{1+3q}\Big)\,k^2\,-\,\frac{4(5-q)}{1+3q}\,k\,-\,\frac{8(1-q)}{1+3q}\,=\,0,
\eee

\ni and again, if we make $q=1$ in (\ref{cc6}) we have that

\bee
\label{cc7}
k\,\Big( k^2 \,+\,\frac 83 k\,-4\, \Big) = 0,
\eee

\ni and the roots are

\baa
\label{cc8}
k_1 &=& 0,\nonumber \\
k_2 &=& \frac 23 (-2 + \sqrt{13}), \nonumber \\
k_3 &=& -\, \frac 23 ( 2 + \sqrt{13}),
\eaa

\ni and the root $k=0$ is the one.  From these results we can see that the assumption in Eq. (\ref{cc1}) is a new relation between $k$ and $q$ (concerning Verlinde's scenario) that confirms $k=q-1$.

With these results we can  write the expressions using KS we have to use that $G \longrightarrow f_{NE} G$ where in general terms we have that

\bee
\label{cc8-1}
f_{NE}= \left\{
\begin{array}{ll}
\displaystyle f(q) \qquad  \mbox{for} \quad  TT \,\,,\\
\displaystyle f(\kappa) \qquad  \mbox{for} \quad KS \,\,.
\end{array}
\right.
\eee

\ni In the near future we will use this compact form to relate the gravitational constants.

\section{Negative Heat Capacity in self-gravitating system and Gravothermal Collapse}
\label{GC}
\renewcommand{\theequation}{5.\arabic{equation}}
\setcounter{equation}{0}

The gravothermal instability described by Antonov  \cite{antonov} has an important role in gravitational thermodynamics  \cite{lw,padmanabhan,chavanis}, namely, the thermodynamics of self-gravitating systems.
A self-gravitating system is a system where its constituents interact with each other trough gravitational forces. 

The discovery of Antonov analyzes a self-gravitating system confined inside a uniform quadratic box.  In this scenario, no maximum entropy state can exist below a critical energy.  After that, this result for energy below this critical value was obtained, where the system would collapse and overheat  \cite{lynden-bell}.  This is the ``gravothermal catastrophe" or ``Antonov's instability."  It is related to the very specific property of self-gravitating systems to have negative specific heats \cite{le}. 
In other words, when discussed in the context of micro-canonical ensemble a self-gravitating system exhibits a negative heat capacity  \cite{padmanabhan}. The negativeness of the heat capacity has been extensively studied in many astrophysical phenomena. For example, we can mention globular cluster  \cite{LBW} and black hole thermodynamics  \cite{Bek2}.

The gravothermal catastrophe plays a relevant role in the evolution of globular clusters.  Concerning dense clusters of compact stars, the gravothermal catastrophe can originate the formation of supermassive black holes.  And statistical mechanics is also relevant for collisionless self-gravitating systems  \cite{chavanis2}.

\subsection{Negative heat capacity}

A negative heat capacity $C_V$ can be explained through the virial theorem for inverse square forces concerning bounded systems.  It connects the average kinetic energy (K) with the average potential energy (V)

\bee
\label{aaaa}
<K>\,=\,- \frac 12 <V>
\eee

\ni and the total energy is given by

\bee
\label{aaab}
E\,=\, <K>\,+\,<V>
\eee

\ni and using (\ref{aaaa}) one has that

\bee
\label{aaac}
E\,=\,-<K>\,\,.
\eee

\ni For moving particles we have that $K= \frac 32 N K_B T$ and consequently the specific heat is negative

\bee
\label{aaad}
C_V\,=\,\frac{dE}{dT}\,=\,-\,\frac 32 \,N K_B\,\,,
\eee

\ni where, for systems with negative $C_V$ that enters in contact with large thermal reservoir, will have fluctuations.   These fluctuations add energy and make its transient temperature lower.   It causes inward heat flow which will head it to even lower temperatures.   Consequently, $-\,C_V$ systems cannot have thermal equilibrium \cite{glass}.

The self-gravitating system, according to the virial theorem, has the total energy $E=-K$ where $K$ is the kinetic energy, $K=\frac{3}{2} N k_B T$. Therefore, the heat capacity of the system is 

\begin{eqnarray}
C_V=\frac{dE}{dT}=-\frac{3}{2} N k_B,
\end{eqnarray}

\ni which is a negative quantity. This negative value indicates an unusual behavior in which the system grows hotter.
Silva and Alcaniz \cite{SA}, by using the kinetic foundations of TT, found a relation for the heat capacity as

\begin{eqnarray}
\label{cvt}
C_V^q=-\frac{3}{5-3q} N k_B,
\end{eqnarray}

\ni which reduces to the Maxwellian limit $C_V=-\,\frac 32 N k_B $ for $q=1$. The negative value of Eq.(\ref{cvt}) shows that the particular $q$-dependence of $C_V$ leads to the classical gravothermal catastrophe ($C_V < 0$)  for the range $0\leq q < 5/3$.

In the $\kappa$-formalism, 
$\kappa$ is the equipartition formula in a 3-dimensional space given by Eq. (\ref{keq}) and  the heat capacity ($C_V=\frac{dE}{dT}$) in the $\kappa$-formalism is given by

\begin{eqnarray}
\label{cvk}
C_V^\kappa=-\frac{3}{2} N \;\; \frac{(1-\frac{1}{2} \kappa)^2\,(1+\frac{3}{2} \kappa)}{(1-\frac{3}{2} \kappa)^2\,(1+\frac{1}{2} k) \, 2\kappa}\; \;
\frac{\Gamma{(\frac{1}{2\kappa}-\frac{3}{4})\,\Gamma{(\frac{1}{2\kappa}+\frac{1}{4})}}}{\Gamma{(\frac{1}{2\kappa}+\frac{3}{4})\,\Gamma{(\frac{1}{2\kappa}-\frac{1}{4})}}}\;\;\; k_B.\\ \nonumber
\end{eqnarray} 

Since

\bee
\label{cc2}
\lim \limits_{k \to 0}
\frac{\Gamma{(\frac{1}{2\kappa}-\frac{3}{4})\,\Gamma{(\frac{1}{2\kappa}+\frac{1}{4})}}}{\Gamma{(\frac{1}{2\kappa}+\frac{3}{4})\,\Gamma{(\frac{1}{2\kappa}-\frac{1}{4})}}}\,=\,2\,k,
\eee

\ni the Maxwellian limit, $C_V=-\,\frac 32 N k_B$, is determined for $\kappa=0$.  The normalized heat capacities of Tsallis' formalism, $C_V^T/N k_B$, (dashed line), and Kaniadakis' formalism, $C_V^\kappa/N k_B$, (solid line), as functions of $\kappa$, are plotted in Figure 1.
As we can see, the heat capacity in the $\kappa$ formalism is negative in the range $0\leq \kappa < 2/3$.  Thus, similarly to TT formalism, the heat capacity in the $\kappa$-formalism also leads to the classical gravothermal catastrophe ($C_V<0$) in the range $0\leq \kappa < 2/3$.
It is worth mentioning that we have used the relation $\kappa=q-1$ which will be discussed in section \ref{mgc}.

\begin{figure}[ih]
\includegraphics[scale=1.1,angle=0]{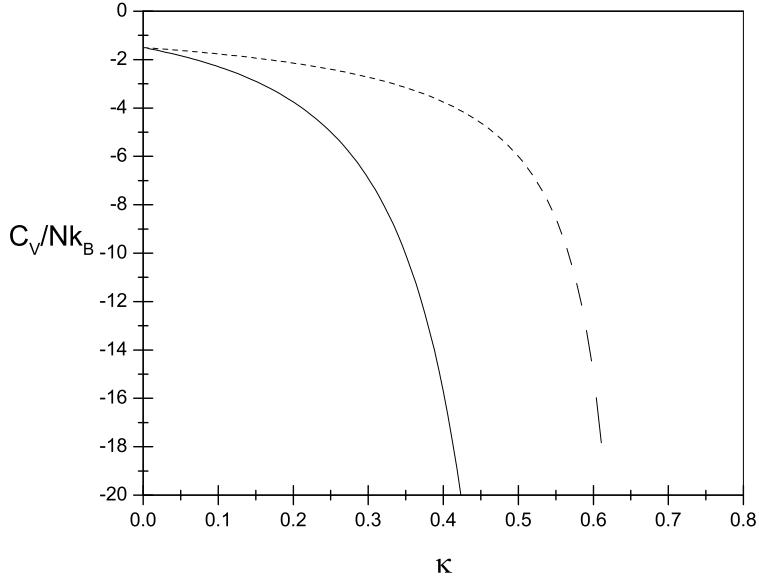}
\caption{The normalized heat capacities for Tsallis (dashed line) and Kaniadakis formalisms (solid line) as functions of $\kappa$ ($\kappa=q-1$).}
\end{figure}

\subsection{Gravothermal catastrophe}

With this statistical mechanics motivation in mind, let us consider a self-gravitating fluid sphere at constant total energy $E$ and volume $V$.  In the case of an instability we have a gravothermal catastrophe and the total energy $E$ of the fluid is

\bee
\label{bb1}
E\,=\, \frac 32 \, N k_B T \,+\, \frac 12 \int \rho(\vec{r}) \Phi(\vec{r}) d^3 r,
\eee

\ni where the second term represents the gravitational contributions.  In the light of the results obtained in Section \ref{GC}, we can write Eq. (\ref{bb1}) as

\bee
\label{bb2}
E\,=\, \frac 32 \, N f(\kappa) k_B T \,+\, \frac 12 \int \rho(\vec{r}) \Phi(\vec{r}) d^3 r,
\eee

\ni and using Kaniadakis equipartition theorem we can make $G \longrightarrow f(\kappa)G_k$  in (\ref{bb2}) we have the gravitational potential becomes

\bee
\label{bb3}
\Phi(\vec{r})\,=\,-\; G_k \int d^3 r' \frac{\rho(\vec{r}')}{|\vec{r}-\vec{r}'|}\,\,.
\eee

\ni Substituting this potential into Eq. (\ref{bb2}) we have that the total energy of the fluid in the Kaniadakis formalism is given by

\bee
\label{bb4}
E\,=\, \frac 32 \, N f(\kappa) k_B T \,-\, \frac 12 \frac{G}{f(\kappa)} \int \frac{\rho(\vec{r})\,\rho(\vec{r}')}{|\vec{r} - \vec{r}'|} d^3 r d^3 r' \,\,.
\eee

\ni where we can clearly that the $\kappa$-formalism function $f(\kappa)$ affect both terms in this equations in a way that we cannot write something like 
$f(\kappa)\,E_{\kappa}$.

From (\ref{bb4}) we can find an expression for the temperature as a function of the density distribution at a constant total energy.  So, in Eq. (\ref{bb4}), if $E=\mbox{constant},\: T=T_0 + T_1$ and $\rho=\rho_0 + \rho_1$, where the indice $0$ means ``unperturbed," we have the expression for $T_1$ as being  \cite{SSL}

\bee
\label{bb5}
T_1 \,=\, -\,\frac{\int \rho_1 (\vec{r}) \Phi_0 (\vec{r}) d^3 r}{\frac 32 N f(\kappa) k_B},
\eee

\ni where $\Phi_0$ is the gravitational potential of the unperturbed density distribution ($\rho_0)$. Consequently, we can directly see that the perturbative temperature is a consequence of KS analysis.

\section{Relative deviations of the modified gravitational constants}
\label{pm2}
\renewcommand{\theequation}{6.\arabic{equation}}
\setcounter{equation}{0}

In order to study the behavior of the modified gravitational constants, Eqs. (\ref{S}) and (\ref{Gk}), when we change the $\kappa$ parameter, we define relative deviations functions as

\begin{eqnarray}
\label{errorq}
\delta G_q=\frac{G-G_q}{G}=\frac{3}{2}\kappa,
\end{eqnarray}
 
\begin{eqnarray}
\label{errork}
\delta G_\kappa=\frac{G-G_{\kappa}}{G},
\end{eqnarray}
where we have used Eqs. (\ref{S}) and (\ref{rel}) in (\ref{errorq}).
In Figure 4 we plot the relative deviations, Eqs. (\ref{errorq}) and (\ref{errork}), as functions of the deformation parameter $\kappa$.
\begin{figure}[t]
\includegraphics[scale=1.1,angle=0]{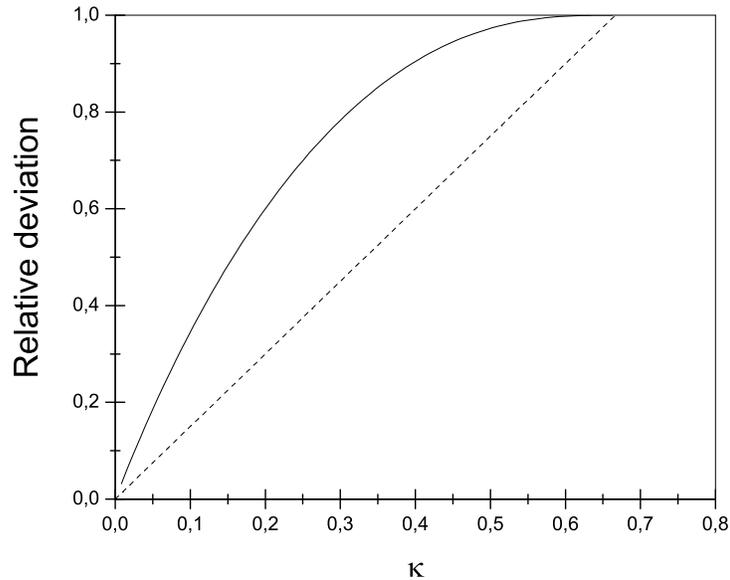}
\caption{The relative deviations of the modified gravitational constants obtained by Tsallis(dashed line), Eq.(\ref{errorq}), and Kaniadakis statistics(solid line), Eq.(\ref{errork}), as functions of $\kappa$.}
\end{figure}
Then, from Figure 4 we can observe that for the same $\kappa$ parameter the value of (\ref{errork}) is greater than (\ref{errorq}) except for $\kappa=0$ and $\kappa=2/3$.






\section{Conclusions and perspectives}

In this work we have derived the equipartition theorem for the Kaniadakis statistics. Our procedure is based on the generalized Maxwellian formulation for $\kappa$-statistics. In the first application, we have analyzed the self-gravitating system in the context of Kaniadakis' formalism, and likewise, in TT formalism we have also obtained the gravothermal catastrophe for all $\kappa$ in the range $0\leq \kappa <\frac{2}{3}$. 

In the second one, we have discussed the Kaniadakis equipartition law of energy in the framework of Verlinde's entropic formalism. We have derived a gravitational constant as a function of $\kappa$ parameter. Comparing with the gravitational constant obtained by TT formalism, we have verified that the difference between both gravitational constants is not significant (figure 3). This result indicates that, in the context of the entropic framework, Tsallis' formalism is more suitable when compared with $\kappa$ statistics because the algebraic expression of equipartition theorem, and consequently the modified gravitational constant obtained by Tsallis' formalism is more simple. Last, the relative deviations of the modified gravitational constants were studied.

As a perspective we can  analyze the formalism developed here, i.e., the connection and the extension of the statistical frameworks, as well as the properties of the modified $G$, to treat high scales cosmological structures.   It is an ongoing research and will be published elsewhere.

\section{Acknowledgments}

\ni   E.M.C.A. thanks CNPq (Conselho Nacional de Desenvolvimento Cient\' ifico e Tecnol\'ogico), for partial financial support.
R.C.N. acknowledges financial support from CAPES ( Coordena\c{c}\~ao de Aperfei\c{c}oamento de Pessoal de N\' ivel Superior), Scholarship Box 13222/13-9.   CNPq and CAPES are Brazilian scientific support agencies.

\end{document}